# Scaling to 1024 software processes and hardware cores of the distributed simulation of a spiking neural network including up to 20G synapses

Elena Pastorelli[1], Pier Stanislao Paolucci[1*], Roberto Ammendola[2], Andrea Biagioni[1], Ottorino Frezza[1], Francesca Lo Cicero[1], Alessandro Lonardo[1], Michele Martinelli[1], Francesco Simula[1], Piero Vicini[1]

[1]INFN Roma "Sapienza", Italy
[2]INFN Roma "Tor Vergata", Italy

[*]Corresponding author: Pier Stanislao Paolucci, E-mail pier.paolucci@roma1.infn.it

**ABSTRACT**

This short report describes the scaling, up to 1024 software processes and hardware cores, of a distributed simulator of plastic spiking neural networks (DPSNN). A previous report demonstrated good scalability of the simulator up to 128 processes. Herein we extend the speed-up measurements and strong and weak scaling analysis of the simulator to the range between 1 and 1024 software processes and hardware cores. We simulated two-dimensional grids of cortical columns including up to ~20G synapses connecting ~11M neurons. The neural network was distributed over a set of MPI processes and the simulations were run on a server platform composed of up to 64 dual-socket nodes, each socket equipped with Intel Xeon Haswell E5-2630 v3 processors (8 cores @ 2.40GHz clock). All nodes are interconnected through an InfiniBand network. The DPSNN simulator has been developed by INFN in the framework of EURETILE and CORTICONIC European FET projects and will be used by the WaveScalES team in the framework of the Human Brain Project (HBP), SubProject3 – Cognitive and Systems Neuroscience. This report lays the groundwork for a more thorough comparison with the neuronal simulation tool NEST.

## 1. Introduction

The Distributed simulator of Plastic Spiking Neural Networks (DPSNN) is a mixed time and event-driven spiking neural network simulator implementing synaptic spike-timing dependent plasticity, designed to be natively distributed and parallel[1].

In this study, we present the scalability of the DPSNN up to 1024 software processes (and hardware cores) and large neural networks including up to 20.4G equivalent synapses (of which 14.2G internal synapses and the remainder simulating an external stimulus). The capability to scale up to such a size of a problem allows simulating entire cortical areas, enabling the study of slow waves in large scale cortical fields [1][2][3], supporting the needs of the CORTICONIC project [4].

In this report, we present measures for two-dimensional grids of up to 96x96 cortical columns, each one including 1240 point neurons (no dendritic tree is represented) and projecting ~1240 synapses per neuron. Assuming a 100µm reticular spacing between neighboring columns, the larger simulated problem size corresponds to the simulation of nearly a square centimeter of cerebral cortex, represented by 11.4M neurons and 14.2G recurrent synapses. Even if full match with biological circuit architecture is not yet a requirement for the level of abstraction adopted up to now by DPSNN users, it is worth noting that the number of neurons per unit area is already near to what needed by biologically accurate simulations. However, the number of synapses per neuron should grow to ~10000 [13][14] for additional biological accuracy.

---

[1] Initially, the DPSNN simulator has been developed as a mini-application benchmark to be used to characterize parallel software and hardware architectures and to drive the development of future generations of neural simulation systems [5][6][7], in the framework of the EURETILE EU project (2010-2014) [8]. Later on, during the CORTICONIC EU project (2013-2016) [4], it has been improved including some of the main features of the Perseo simulator – a scalar (not parallel, nor distributed) event-driven simulator of spiking neural networks based on the studies of P. Del Giudice and M. Mattia [9] – with the aim to build an efficient tool for large scale cortical simulations in support of in-vivo and in-vitro experimental results. In a near future (starting from April 2016), DPSNN will be used for the study of the Slow Wave Activity in the framework of WaveScalES, a part of the SP3 subproject of the Human Brain Project (HBP) [10]. Moreover, it will also be used for a comparison with the well-known NEST simulator [11][12], in order to give a contribution to the improvement of the neural network simulation platforms in the HBP context.

## 2. Methods

A detailed description of the parallelization techniques and internal architecture of the DPSNN together with its strong and weak scaling behavior on a range of neural configurations spanning from larger networks, composed of 6.6G synapses and 32.8M neurons, down to smaller configurations composed of 200K synapses and 1K neurons is reported in Paolucci et al., 2013 [5][7]. In that work, between 1 and 128 software processes (and hardware cores) were used for distributed execution. Hereafter, the reader will find a very brief summary of the simulator architecture, while the focus of this document is on performance analysis and scaling measurements, in the specific context set by the requirements of the CORTICONIC project [4].

DPSNN is a spiking neural network simulator coded as a network of C++ processes, designed to be easily interfaced to both MPI and custom Software/Hardware Communication Interfaces. The report [5] describes the behavior of a simulated neural network composed of a mixture of 80% Regular Spiking (RS) Excitatory, 20% Fast Spiking (FS) Inhibitory Izhikevich point-like neurons, with STDP (Spike-Timing Dependent Plasticity) evolution of synaptic weights.

The measurements reported herein are relative to columns of Leaky Integrate and Fire (LIF) neurons with spike-frequency adaptation (SFA) due to calcium- and sodium-dependent after-hyperpolarization (AHP) currents [15][16]; in these measures, synaptic plasticity was disabled, because plasticity is not a requisite in the framework of the CORTICONIC project. The neural columns, composed of 80% excitatory and 20% inhibitory point-like neurons, have been arranged in bi-dimensional grids, with decreasing synaptic connectivity according to the distance between column pairs. For these measurements, we selected a grid step $\alpha \sim 100$ µm. Connectivity can be varied according to the simulation needs, leading to configurations with different numbers of synapses per neuron. In the reported case, the local connectivity, i.e. the probability of connection between neurons placed in a same column, is set to 80% while the lateral (or remote) connectivity, i.e. the probability of connection among neurons located in different columns, decreases with the distance proportionally to $A \cdot \exp(-r^2/2\alpha^2)$, with A=0.05. The remote connectivity function is similar to that adopted by Potjans and Diesmann, 2014 [14], although with different A and $\alpha$ parameters. A cut-off has been inserted during the generation of remote synapses, restricting the projections to the subset of columns with connection probability no lesser than 1/1000; in this case, this translates to a centered 7x7 stencil around each column in the grid. The number of neurons per column was fixed to 1240. The connectivity just described produces between 1239 and 1245 synapses/neuron. About 990 synapses are projected toward the same column (80% probability) while the remainder is the set of remote synapses.In addition to the recurrent synapses, the system simulates also a number of external synapses: they represent afferent (thalamo-)cortical currents coming from areas external to the network. These additional synapses are schematized as additional excitatory inputs to each neuron, $C_{ext}$. In our model $C_{ext}$ has been fixed to 540. The recurrent synapses plus the external synapses gives the number of total synapses afferent to a neuron, referred to as total equivalent synapses.In order to study the scalability of the DPSNN, measurements were taken on different problem sizes obtained varying the dimension of the grid of columns and, for each problem, distributing it over a span of MPI processes. We selected three grid sizes which, assuming columnar distance of 100 µm, can already be considered representative of interesting biological dimensions: 24x24, 48x48, 96x96 columns. Table 1 summarizes the main figures of the different systems used in our simulations.

| Grid | Number of Columns | Number of Neurons | Number of Recurrent Synapses | Total Equiv. Syn (external included) |
|---|---|---|---|---|
| **24x24** | 576 | 0.7M | 0.9G | 1.2G |
| **48x48** | 2304 | 2.9M | 3.5G | 5.0G |
| **96x96** | 9216 | 11.4M | 14.2G | 20.4G |

Table 1 - Problem sizes used in the scaling measurements

The server platform used to run the simulations described herein is GALILEO [17], a cluster of 516 IBM nodes provided by the CINECA consortium. Each 16-core dual-socket computational node contains two Intel Xeon Haswell 8-core E5-2630 v3 processors, clocked @2.40GHz. All nodes are interconnected by an InfiniBand network. The platform configuration did not permit hyper-treading, therefore for these measures the number of cores matches how many MPI processes were launched at each execution.

### 3. Results

This report is about the strong and weak scaling behavior of the DPSNN up to 1024 software processes/hardware cores on neural networks including up to 20.4GSyn interconnecting a 96x96 grid of cortical columns. It extends the explored range of DPSNN scaling that was already tested up to 128 software processes/hardware cores [5] for networks composed of up to 6.6G synapses and 32.8M neurons down to smaller networks including 200K synapses and 1K neurons.

Each neuron receives a total of synaptic events from both recurrent synapses (i.e. internal to the network) and external synapses. Strong scaling measures are expressed as elapsed time per total number of synaptic events. Accordingly, the measures used in weak scaling are expressed in units of elapsed time per synaptic event per core.

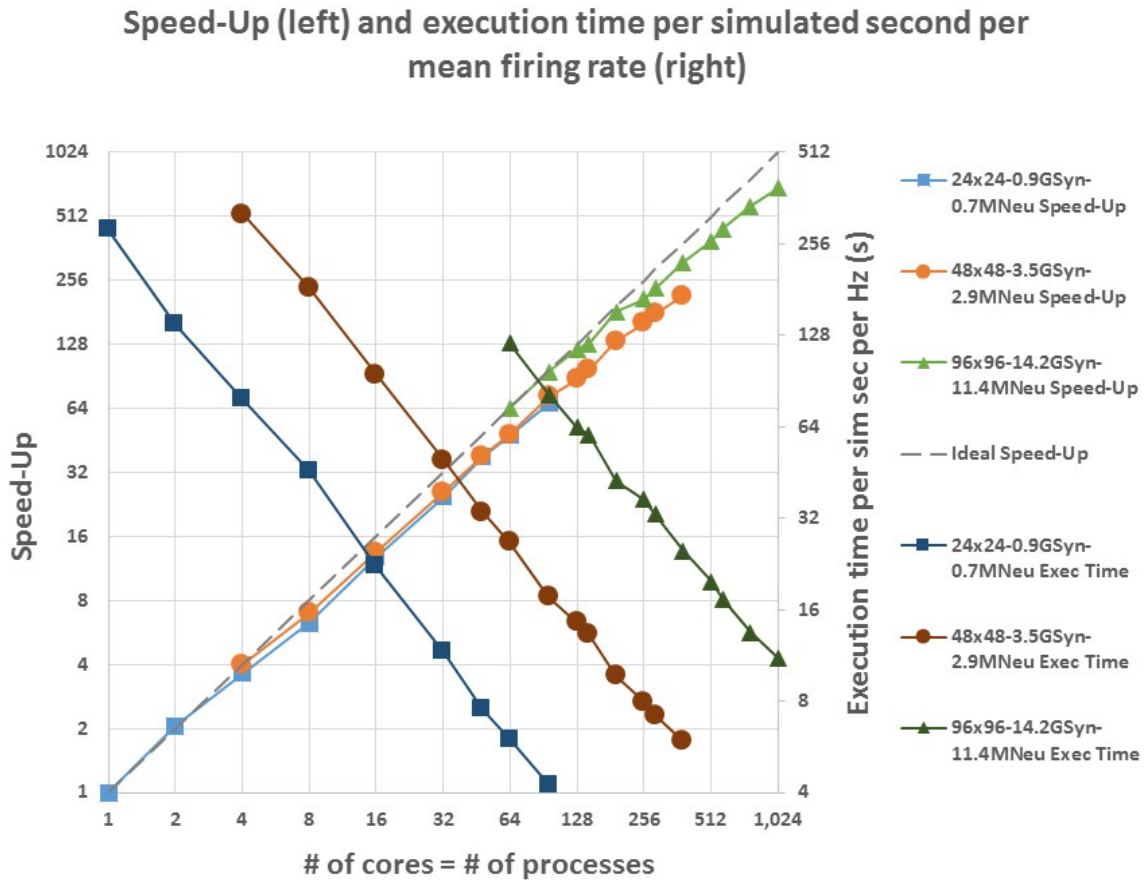

Figure 1 – Speed-Up and Execution Time

**Execution time and speed-up**

Figure 1 reports the speed-up and execution times for 3 different configurations (24x24-0.9GSyn-0.7MNeu, 48x48-3.5GSyn-2.9MNeu and 96x96-14.2GSyn-11.4MNeu grids) as a function of the number of cores assigned to the problem. Due to the particular configuration of the used resources, the number of cores corresponds also to the number of processes launched at each execution. The speed-up curves (left vertical axis), move from bottom left to upper right. The black dashed line represents the ideal scaling. The measured execution times are the three descending curves (right vertical axis).

**Strong scaling**

The strong scaling of the DPSNN is in Figure 2. The values represent how the execution time per synaptic event changes when the number of cores assigned to the problem is varied. In the ideal case (black dot line in the picture), doubling the used resources, execution time should halve. In our measures, the time needed to simulate the 24x24 grid (with 0.9G recurrent synapses and 1.2G total equivalent synapses) scales down from $2.75 \times 10^{-7}$ seconds per synaptic event, using 1 single core, to $4.09 \times 10^{-9}$ seconds per event using 96 cores. The actual speed-up is 67.3, loosing ~30% compared to the ideal speed-up (96). The speed-up for the 48x48 grid (3.5G recurrent, 5G equivalent synapses) is 54.2 while the hardware resources increase by a factor 96. For the 96x96 grid (14.2G recurrent/ 20.4G equivalent synapses) the speed-up is 10.8 (16 would be the ideal).

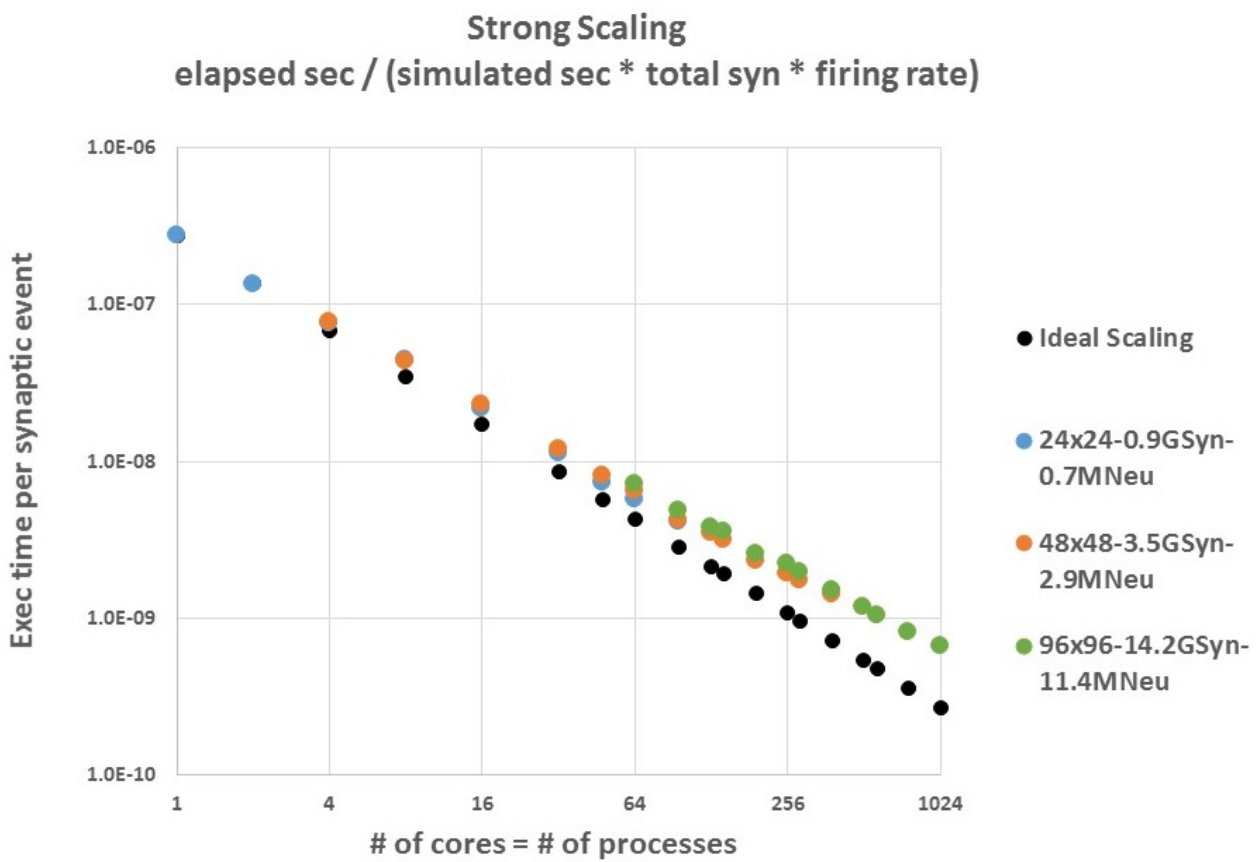

Figure 2 – Strong Scaling. The measures are expressed in elapsed time per synaptic event, where the synaptic event represents each excitatory and inhibitory synaptic current reaching the neuron both from recurrent and from external synapses.

**Weak scaling**

The values reported in the diagram of Figure 3 represent the scaling when the problem size assigned to each core is kept constant and the total problem size is increased proportionally to the number of hardware cores assigned to the solution (horizontal axis). For a fixed problem size/core, an ideal code, should exhibit constant execution time. Ideal weak scaling should produce horizontal lines for each problem size/core. If normalized by the load per core, the lines corresponding to different loads/core should overlap.

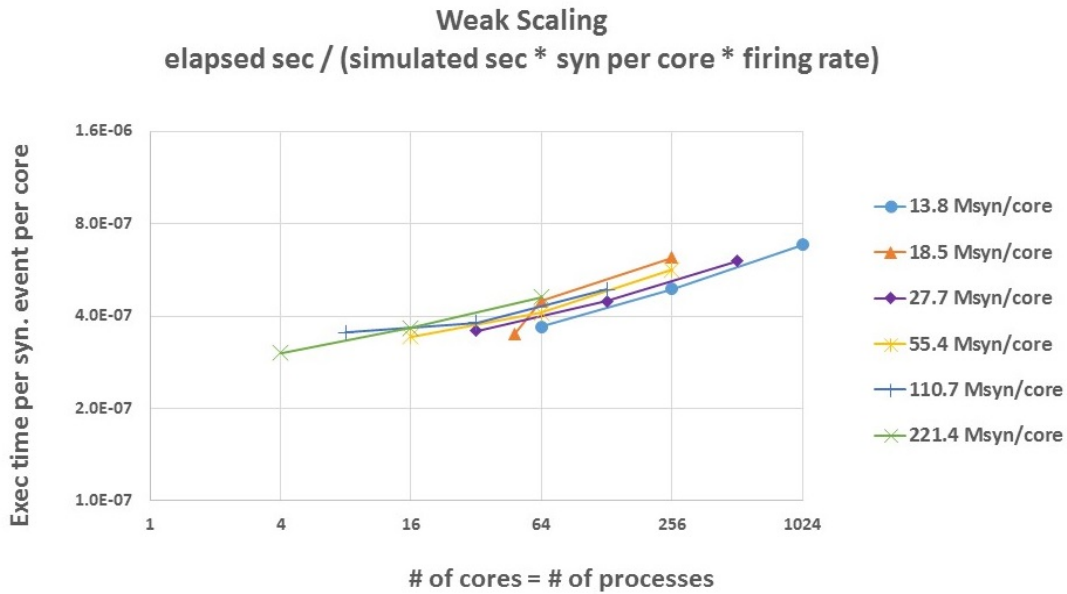

Figure 3 – Weak Scaling

**Memory**

The memory used to run the three configurations was gauged. In Figure 4 the bytes per synapse are reported for the three selected grids varying the number of software processes/hardware cores. As a result, the number of bytes/synapse used to simulate the three systems (24x24, 48x48 and 96x96 grids) is between 25.9 and 34.4 bytes/synapse.

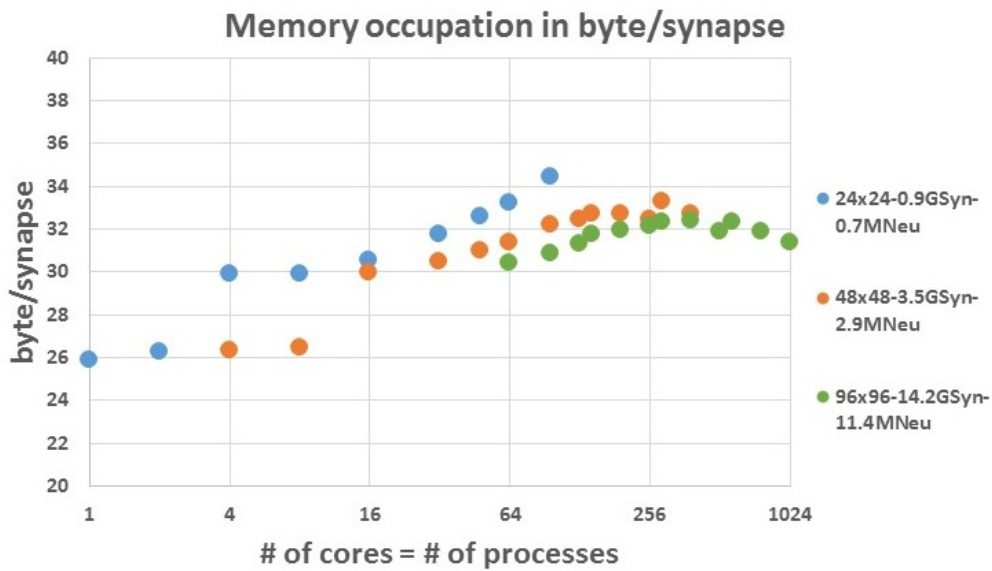

Figure 4 - Used memory per synapse

## 4. Discussion

This paper presents an extension to 1024 software processes (run on 1024 hardware cores) of the scaling measurements of the DPSNN previously reported in [5][7] for up to 128 processes and hardware cores. We demonstrated the ability to simulate a grid of 96x96 neural columns, containing a total of 11.4M LIF neurons with spike-frequency adaptation, and representing 20.4G equivalent synapses. Assuming a cortical column spacing of about 100 μm, this problem size could represent a cortical tissue area of about one square centimeter. We proved that DPSNN is able to simulate this kind of problem size about eleven times slower than real-time on a 1024 core execution platform, with a memory occupation below 34.4 byte-syn. In addition, thanks to the contribution of the ISS (Istituto Superiore di Sanità, Rome, Italy), the distributed simulator DPSNN incorporates the main functionalities offered by the scalar simulator Perseo [9]. In the framework of the CORTICONIC project, the capability to scale to such a problem size, joined with integration of additional functionalities, enables the study of slow waves in large scale cortical fields [1][2][3].

## 5. Acknowledgements

The DPSNN simulator development has been partially funded by the EURETILE European project and by the CORTICONIC European project. We also acknowledge the contribution of Paolo Del Giudice and Maurizio Mattia, in the framework of CORTICONIC. Further, we thank Leonardo Cosmai for his support in running simulations on the Galileo supercomputer at CINECA.